# A Self-Assessing Compilation Based Search Approach for Analytical Research and Data Retrieval

Ananth Goyal [Sole Author]
Dougherty Valley High School; San Ramon, USA [Institution]

**Abstract**—While meta-analytic research is performed, it becomes time-consuming to filter through the sheer amount of sources made available by individual databases and search engines and therefore degrades the specificity of source analysis. This study sought to predict the feasibility of a research-oriented searching algorithm across all topics and a search technique to combat flaws in dealing with large datasets by automating three key components of meta-analysis: a query-based search associated with the intended research topic, selecting given sources and determining their relevance to the original query, and extracting applicable information including excerpts and citations. The algorithm was evaluated using 5 key historical topics, and results were broken down into 4 categories—the total number of relevant sources retrieved, the algorithm's efficiency given a particular search, the total time it takes to finish a complete cycle, and the quality of the extracted sources when compared to results from current searching methods. Although results differed through several searches, on average, the program collected a total of 126 sources per search with an average efficiency of 19.55 sources per second which, when compared and qualitatively evaluated for definitive results, indicates that an algorithm developed across all subject areas will make progress in future research methods.

## 1 INTRODUCTION

A searching algorithm is any computational algorithm finding a specified attribute within a collection of data or dataset array (data repositories, data archives, etc.)[1]. The term 'search engine' is used in the context of this study, because it considers relevance and association with a query rather than direct equivalence between two data components found on the web [2]. The most widely used search engine is Google [3], which conducts about 92% of all searches worldwide, and is by far the most accurate and powerful search engine [4][5]. Although Google, and its subdivision Google Scholar are holistically superior to other search systems, their broad scope frequently compromises the quality of results with a preference for high speed and lower runtimes [6], still requiring the user to assess each result making it impractical for focused needs. Additionally, most reliable information that is provided by research-oriented search engines, such as Google Scholar require payment and have biases [7], when sources of equivalent, credible, and impartial material in the public domain are available.

This paper demonstrates a newly proposed algorithmic concept for countering these predicaments and for predicting whether its mechanism can be extended to all searching systems, specifically when conducting research or organizing large quantities of data. The algorithm can be broken down into two proposed novel categories:

*Multitudinous Database Search (MDS)* uses several topic-related databases and their built-in searching mechanisms to increase the number of sources by which information is being extracted, ensuring that the scope of the search is optimal for the focus of the query.

*Source Analysis and Extraction Algorithm (SAEA)* filters through several search results deciding if a specific source should be used and what specific components of those sources should be extracted (quotes, excerpts, citations, images, data, etc. )

The rest of the paper is structured as follows: The second section *Prior Work* discusses current algorithms referenced in this paper. Section three *Methods* elaborates on the proposed algorithm and its mechanics. All results and data collected from experimental testing will be compared to Google Scholar and are found in the *Results* section followed by *Conclusions*.

## 2 RELATED WORK

This section discusses extant searching techniques/methods germane to the proposed algorithm.

A Self-Assessing Compilation Based Search Approach for Analytical Research and Data Retrieval

**On-Site Searching Methods**

Most searching systems within a site or database primarily operate by using the frequency of keywords to determine the relevance of that source or webpage to the query. All the words in a given search query are of equal importance and relevance [8], making it imperative for the user to refrain from including any conjunctions and filler words to the query that would produce a coherent sentence—diverting the focus of the search; to streamline the number of results—thereby improving—the accuracy of the results, searching with only keywords and topics is suggested. Repeated searching within every sub-page is avoided by pre-indexing information on page content and their associated topics within a dataset, making the process for displaying relevant links efficient [9][10].

The use of Proximity Searching (PS) is one particular approach to increase the accuracy of on-site search results. PS focuses on the nature of certain keywords, rather than their frequency [11]. Through using predetermined object relationships, results will contain goal information even when it is not directly related to the query [12]. Although Google search is not an on-site searching system, it uses a PS-based algorithm to retrieve tangentially related results when not explicitly specified in the query [13]. For example the word "thriller", in Google's indexing database is in close proximity to Michael Jackson who is renowned for performing a song with the same name.

**Google PageRank**

First Pioneered by Google, PageRank (PR) was designed to be a method of ranking search results by popularity and link information [14]. The PR value of a webpage, node, or link is the page's calculated ranking when compared with other results given a search query [15][16]. The primary aim of this organizational technique is to allow the user to obtain results that are mostly referenced in other sites, since more source references implies greater relevance and significance for the search query [17]. The PR formula is shown below:

$$PR(u) = (1-d) + d\sum_{i=1}^{n}\left(\frac{PR(i)}{L(i)}\right)$$

The letter *'u'* represents any given page prior to calculating its PR value while *'L'* denotes the number of outbound links referenced. The letter *'i'* represents a referenced web page used to calculate the target PR of page *u*. A key aspect of the PR algorithm is the damping factor (denoted as *'d'*) which is the constant reduction in probability for the user to click on any further links after having already analyzed the information found in prior ones [18][19]. For example, if page 'A' has a higher PR score than others, then it will rank first on the PR result page; however, since the following links are lower in reference popularity the user may refrain from opening new links after only a few, making it critical to employ a damping factor.

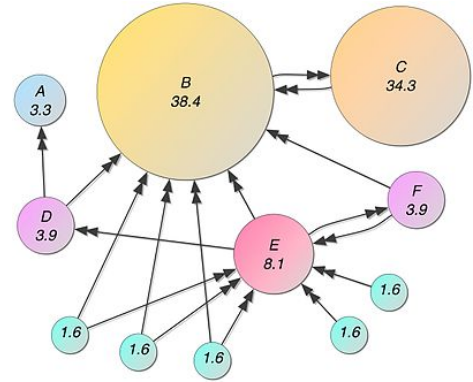

**Fig. 1:** Depicts variability in PR scores from factors including page importance, relevance, outbound links, and damping factor [20]. In this scenario, page 'B' has the highest PR score (38.4), likely making it the first link to appear after a Google search.

The inherent nature for researchers and students to avoid lower ranked PR links is inevitable, thus the proposed solution (explained in "Methods") addresses this problem by automating data extraction from all relevant sources listed.

## 3 METHODS

Each database used in this study was a public domain historical primary source bank, each with different interfaces and mechanics, requiring slightly altered web-crawling methods to optimize performance. *MDS* was applied using Proximity Searching techniques from the integrated search system of the databases to generate a result page of all links related to content grouped by relevance and popularity (PR). Then, *SAEA* was used to analyze and accumulate the relevant data from every source to be displayed on the final page, which consisted of the primary source document (speech, article, image, map, etc.) and citation, all organized by their respective relevance scores.

The historical databases used in this study were DocsTeach National Archives , Yale Avalon Project (YA), EyeWitness to History (EW), The Ancient Encyclopedia (AE), and the John Carter Brown Library (JCB). To produce the best results and comply with the variability of searching mechanics, a separate class has to be built and implemented for each database. Each class consists of two key methods, one designed for *MDS* and one built for *SAEA*. The

A Self-Assessing Compilation Based Search Approach for Analytical Research and Data Retrieval

retrieved information was all compiled and displayed on a local web file.

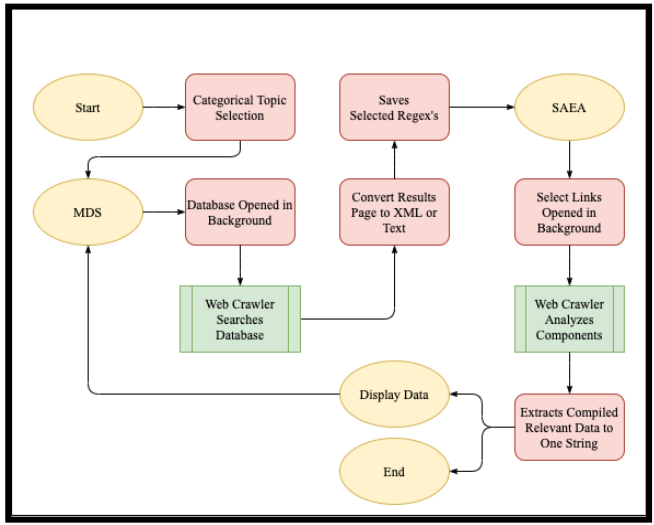

**Fig. 2:** Depicts the whole algorithmic process from search to display including the most notable components..

**Multitudinal Database Searching (MDS) System**

MDS is influenced by on-site searching and its subdivision Proximity Search, as it replicates a single search in a database across several, maximizing the amount of information that could potentially be available to the user. Prior to the submission of the search query, the user focuses the scope of their research by selecting categorical topics that conform to their desired results. This reduces the number of databases the MDS system would have to reference, reducing the time for which the algorithm will operate. After the query is submitted, the web-crawler uses the built-in search mechanisms of the databases by parsing the query into a search format, which is then inputted into the search parameter of the URLs. Some databases use Google's "Custom Search" algorithm to index and retrieve their information [21], in that case, MDS is applied through that mechanism.

The links to each of the articles and sources retrieved from each database are saved into a dataset. The links are separated from the remaining XML or text components by continuously approximating their location using index search function to target the particular regex/regular expression (used to describe a component of a string [22]) for however many links present in the file page of the results. This process occurs simultaneously for every database until every set of resulting URLs are stored into a data set.

**Source Analysis Extraction Algorithm (SAEA)**

For keyword relevance, each stored link is opened and evaluated to filter out any non-applicable components of articles. All relevant information, including the citation which is found by regex approximation, is then stored into a string to be displayed after all the information from the entire database has been collected. SAEA is developed to bypass the PR system for each database/Google result page, as it analyzes more than the first 1-3 resultant pages. Unlike Google/Scholar, instead of simply listing out links with a brief description, SAEA opens the links and compiles what the user intends to extract themselves, automating the entire search process altogether.

The method by which SAEA extracts a component from a source is entirely dependent upon the database being used and the type of data being processed. Different methods and subroutines are developed to properly extract the information, depending on the type of source. If the user selects image files as their target type, then the retrieval method is standard; isolate the image tags found in the XML script. However, if a particular type of information other than an image is being targeted, then a specialized algorithm will be developed for identifying it. These include identifying textual patterns in headings, paragraphs, descriptions, etc., looking at the particular formatting of the text, such as a citation, or identifying where in the source/file the target information will be, if it is repetitive or has a significant pattern. These methods are independently developed for each circumstance.

In order to predict the total number of sources compiled through SAEA at any instant in time, timestamp information from each database containing completion data is used to model and predict *S (number of sources)* as a function of *t* (*time*) using a variant *Interpolation Polynomial Formula* (IPF) which outputs a differentiable function that runs through the set of given data points [23]. Data points collected are represented as *(time, # of sources)*; the total number of databases used is represented by '*n*' as each point being used in IPF is the completion cycle for a single database. The letter '*y*' denotes the output of the data-point, which in this case is the number of sources retrieved from this database. The formula is shown below [24]:

$$S(t) = \sum_{i=1}^{n}(y_i)\left[\prod_{k=1}^{n}\left(\frac{t-t_k}{t_i-t_k}\right)\right]$$

IPF is used to model the results collected from every search query; however, by taking the derivative of the IPF polynomial, a new function is constructed to predict the efficiency or speed of the program at any given moment in time. The efficiency function defined as E(t), estimates the instantaneous speed represented as a data-point of (x~seconds, y~sources per second), revealing patterns of more and less effective algorithmic segments.

The average number of retrieved sources is calculated by using an Average Value Theorem (AVT) which calculates a function's average value over a certain interval [25]. In this case, AVT is used to find the average number of sources retrieved using the S(t) function produced from IPF.

$$A(t) = \frac{1}{t_f - t_i} \int_{t_i}^{t_f} S(t)$$

This method is used to measure an atypical mean, as it factors in the possible inclusion of more databases, allowing the resulting data to predict future implications of source retrieval. It is calculated by using two time components, the initial and final bounds of its compilation process which usually range from 3 to 10.

## 4 RESULTS AND DISCUSSIONS

**Quantitative Results**

The numerical data for the proposed algorithm is divided into 3 categories provided a search query—the number of sources retrieved from each database, the time to complete each cycle, and the modeled source and efficiency function.

The results were collected by using a built in method that records the timestamp information of every database, which includes the number of sources retrieved and the amount of time it took to do so. Table 1 displays the data collected from each group across the various databases when given a sample historical search term from different periods. The first number under each database represents the time it takes to complete a cycle and the number below denotes the number of sources retrieved in that cycle ('S').

| Search Query | EW | YA | AE | JCB | Total |
|---|---|---|---|---|---|
| Christopher Columbus | 2.88 sec 33 S | 3.78 sec 46 S | 4.75 sec 8 S | 5.21 sec 54 S | 5.21 sec 141 S |
| Slave Trade | 5.84 sec 27 S | 4.36 sec 56 S | 5.35 sec 13 S | 3.37 sec 43 S | 5.84 sec 139 S |
| WWI | 7.34 sec 32 S | 5.74 sec 45 S | n/a | 4.18 sec 36 S | 7.34 sec 113 S |
| WWII | 7.28 sec 31 S | 6.35 sec 37 S | n/a | 4.72 sec 41 S | 7.28 sec 109 S |

**Table 1:** The collected time stamp information and IPF generated functions for varying search queries across the test databases (n/a means the search term is not from the target time range of the database)

| Search Query | Source and Efficiency Function of Time |
|---|---|
| Christopher Columbus | $S(t) = 14.516t^3 - 188.549t^2 + 821.01t - 1114.36$<br>$E(t) = 43.548t^2 - 377.098t + 821.01$ |
| Slave Trade | $S(t) = 37.1105t^3 - 507.342t^2 + 2306.13t - 3387.15$<br>$E(t) = 111.3315t^2 - 1014.684t + 2306.13$ |
| WWI | $S(t) = -2.79942t^2 + 56.6164t - 151.744$<br>$E(t) = -5.59884t + 56.6164$ |
| WW2 | $S(t) = 4.15389t^2 - 23.2841t + 58.3592$<br>$E(t) = 8.30778t - 23.2841$ |

**Table 2:** Depicts the collected time stamp information and IPF generated functions for varying search queries across the test databases

The following graph is an example of an IPF generated function for the total number of sources and the efficiency / speed function at an instant of time. The dotted lines represent the data's restricted domain, since that is the region of time in which the given functions operate. Use of IPF enables us to model the performance of the algorithm's functional progress for better comparisons and future modifications.

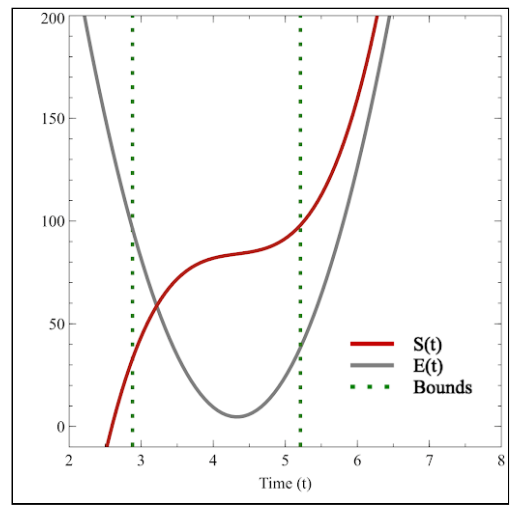

**Fig. 3:** SAEA IPF Graph: "Christopher Columbus"

**Comparative Results of Google Scholar**

Results were compared to test results from searches on Google Scholar (GS). GS is designed so it can deliver thousands to millions of results in less than a second. Identical search queries from the previous tests were used on GS. Although direct results cannot be properly used for comparison, further analysis and predictive modeling make it possible.

| Search Query | Retrieval Time (sec) | Total Number of Sources |
|---|---|---|
| Christopher Columbus | 0.06 | 222,000 |
| Slave Trade | 0.05 | 1,460,000 |
| WWI | 0.03 | 4,610,000 |
| WW2 | 0.05 | 4,780,000 |

**Table 3:** The directly retrieved data after each search on GS. Note: Retrieval time and the number of sources pulled are not indicators of overall effectiveness, but primarily just speed.

The direct retrieval and index speed of GS is significantly superior; however, the necessity for the user to open each link and the lack of automation in source analysis is what SAEA addresses and is what is being compared. GS is not a compilation based search engine and does not self-analyze its results, which are both done by the user. Compilation times are simulated to check its efficacy properly with MDS/SAEA, compilation times will be simulated. The manual compilation time is how long it will take a researcher to extract the intended information from all the selected sources, in this case from each link to the historical primary source material. Only the first page of results (10 sources) should be used, as a PR system indicates the user is unlikely to go any further. Since the secondary purpose of the proposed algorithm is to be cost-effective (free of charge), the cost of all sources available on the first result page (10 sources) will be totaled.

| Search Query | Number of Sources Used | Manual Compilation Time (secs) | Total Cost of Sources Used (USD) |
|---|---|---|---|
| Christopher Columbus | 10 | 414 | $44.64 |
| Slave Trade | 10 | 378 | $56.59 |
| WWI | 10 | 438 | $103.99 |
| WW2 | 10 | 412 | $66.95 |

**Table 3:** Calculated data after each search on GS for MDS/SAEA comparisons.

Results of these methods were contrasted by taking the compilation times average and dividing by the total number of sources used/retrieved. This introduces a new efficiency function, which can be graphically modeled between the time bounds as the derivative/slope of the secant line. On average, it was found after testing that: manual extraction of target information using GS takes about 0.026 sources per second. Comparatively, automatic extraction using the proposed algorithm runs approximately 19.55 sources per second. The purpose of this algorithm is not to replace Google Scholar because, overall, GS has far more capabilities and substantially higher retrieval rates; however, MDS/SAEA would do best when searching for specific components in a paper (images, methods, results), primary source material, citations, excerpts, or any situation where the researcher is frequently searching for a particular component, over the entire source. GS may be viewed as a database itself, and integrated into MDS as if the information extracted came from a single source. This will theoretically extend the reach from which source components can be obtained, and in effect increase the pace and degree to which research is conducted.

## 5 CONCLUSION

In this paper, a new method to enhance analytical research using multiple databases was presented. Upon checking and observing the discrepancies between source quality and efficiency from varying combinations of databases and search queries, it is evident that this method will enhance current research techniques in all subject areas. This research was performed using historical documents only; however, the basic dynamics and functionalities of the algorithmic method would virtually remain identical regardless of the intended subject matter. First, *MDS* is used to find all associated sources and then *SAEA* is used to filter out any irrelevant findings and extract the appropriate components. The final results show the method can retrieve an average of 126 related source results within 4 to 8 seconds, all with an average efficiency of 19.55 at any given moment in time.

**ACKNOWLEDGMENTS**

The completion of this research report and study would not have been possible without the support from Mr. Sudhir Kamath, Mr. Robert Gendron, Professor Jeffery Ullman, and Mrs. Katie MacDougall and I deeply thank them for all their contributions.

**REFERENCES**

[1] A. C. Dalal, "Searching and Sorting Algorithms." [Online]. Available: http://www.cs.carleton.edu/faculty/adalal/teaching/f04/117/notes/searchSort.pdf

[2] S. Brin and L. Page, "The anatomy of a large-scale hypertextual Web search engine," Computer Networks and ISDN Systems, vol. 30, no. 1-7, pp. 107–117, 1998.

A Self-Assessing Compilation Based Search Approach for Analytical Research and Data Retrieval

[3] Patterson, Mark R., Google and Search Engine Market Power (April 27, 2012). Fordham Law Legal Studies Research Paper No. 2047047.

[4] Search Engine Market Share Worldwide," Dec-2019. [Online]. Available: https://gs.statcounter.com/search-engine-market-share.

[5] Ioannis Lianos, Evgenia Motchenkova, MARKET DOMINANCE AND SEARCH QUALITY IN THE SEARCH ENGINE MARKET, Journal of Competition Law & Economics, Volume 9, Issue 2, June 2013, Pages 419–455, https://doi.org/10.1093/joclec/nhs037

[6] Brutlag, J. (2009). Speed matters for Google web search.

[7] Diaz A. (2008) Through the Google Goggles: Sociopolitical Bias in Search Engine Design. In: Spink A., Zimmer M. (eds) Web Search. Information Science and Knowledge Management, vol 14. Springer, Berlin, Heidelberg

[8] "LibGuides: Research Process: Proximity Searching," Proximity Searching - Research Process - LibGuides at Northcentral University, 24-Feb-2020.

[9] R. Shaw, "Understand How On-Site Search Works," Inc.com, 07-Apr-2000. [Online]. Available: https://www.inc.com/articles/2000/04/18342.html.

[10] Marti Hearst, Ame Elliott, Jennifer English, Rashmi Sinha, Kirsten Swearingen, Ka-Ping Yee Communications of the ACM Vol. 45, No. 9 (September 2002), Pages 42-49

[11] "Proximity Operators," 13-Jan-2009. [Online]. Available: https://library.alliant.edu/screens/proximity.pdf.

[12] Goldman, R. and Shivakumar, N. and Venkatasubramanian, S. and Garcia-Molina, H. (1998) Proximity search in databases.

[13] D. M. Russell, "Advanced Search Operators," 23-Aug-2019. [Online]. Available:https://journalismcourses.org/courses/DATA0819/Eng/Google_Advanced_Search_Operators.pdf.

[14] I. Rogers, "The Google Pagerank Algorithm and How It Works." [Online].Available:https://www.cs.princeton.edu/~chazelle/courses/BIB/pagerank.htm.

[15] L. Page, "The PageRank Citation Ranking: Bringing Order to the Web," Stanford InfoLab, Jan. 1998.

[16] Amy N. Langville & Carl D. Meyer (2004) Deeper Inside PageRank, Internet Mathematics, 1:3, 335-380, DOI: 10.1080/15427951.2004.10129091

[17] E. Roberts, "The Google PageRank Algorithm," 09-Nov-2016. [Online].Available:https://web.stanford.edu/class/cs54n/handouts/24-GooglePageRankAlgorithm.pdf.

[18] "The Damping Factor," The Damping (or dampening) Factor, 2004. [Online]. Available: http://www.pagerank.dk/Pagerank-formula/Damping-factor.htm.

[19] Boldi, P., Santini, M., & Vigna, S. (2005, May). PageRank as a function of the damping factor. In Proceedings of the 14th international conference on World Wide Web (pp. 557-566).

[20] Simple example of PageRank. Damping 85%. Area of circles equals PageRank. 2007

[21] Mysen, C. C., Verma, N., & Chen, J. (2011). U.S. Patent No. 8,082,242. Washington, DC: U.S. Patent and Trademark Office..

[22] J. Goyvaerts, Regular Expressions: The Complete Tutorial. Princeton, 2007.

[23] V. Dahiya, "Analysis of Lagrange Interpolation Formula," - International Journal of Innovative Science, Engineering & Technology, vol. 1, no. 10, Dec. 2014.

[24] Archer, Branden and Weisstein, Eric W. "Lagrange Interpolating Polynomial." From MathWorld--A Wolfram Web Resource. http://mathworld.wolfram.com/LagrangeInterpolatingPolynomial.html

[25] P. Dawkins, " Average Function Value," Calculus I - Average Function Value, 30-May-2018. [Online]. Available: http://tutorial.math.lamar.edu/Classes/CalcI/AvgFcnValue.aspx. [Accessed: 07-Mar-2020].